\documentclass[preprint2]{aastex}

\usepackage{fixltx2e} 
\usepackage{amsfonts}
\usepackage{amsmath}
\usepackage{amssymb}

\newcommand{\vbens}{VB04}
\newcommand{\vbdec}{VB06}
\newcommand{\sub}{\textsubscript}
\newcommand{\up}{\textsuperscript}
\newcommand{\angs}{\mbox{\normalfont\AA}}  
\newcommand{\gba}{\emph{g}}
\newcommand{\rba}{\emph{r}}
\newcommand{\iba}{\emph{i}}
\newcommand{\uba}{\emph{u}}
\newcommand{\zba}{\emph{z}}
\newcommand{\dm}{$\Delta m$}
\newcommand{\dmoff}{$\Delta m_{off}$}
\newcommand{\dmc}{\Delta m_{AGN}}
\newcommand{\sn}{S/N}
\newcommand{\sigphot}{$\sigma_{phot}$}

\shorttitle{Optical Variability of AGN in SDSS-DR7}
\shortauthors{Gallastegui-Aizpun \& Sarajedini}

\begin{document}

\title{The Ensemble Optical Variability of Type-1 AGN in the Sloan Digital Sky Survey Data Release 7}

\author{Unai Gallastegui-Aizpun}
\affil{Department of Astronomy, University of Florida, 211 Bryant Space Science Center, Gainesville, FL 32611}
\email{ugallastegui@astro.ufl.edu}

\and

\author{Vicki L. Sarajedini}
\affil{Department of Astronomy, University of Florida, 211 Bryant Space Science Center, Gainesville, FL 32611}
\email{vicki@astro.ufl.edu}


\begin{abstract}

We use a sample of over 5000 active galactic nuclei (AGN) with extended morphologies at $z<0.8$ from the Sloan Digital Sky Survey (SDSS) to study the ensemble optical 
variability as a function of rest-frame time lag and AGN luminosity with the aim of investigating these parameter relationships at lower luminosities than previously studied. 
We compare photometry from imaging data with spectrophotometry obtained weeks to years later in the Sloan \gba, \rba, and \iba~bands. We employ quasar and galaxy eigenspectra 
fitting to separate the AGN and host galaxy components. A strong correlation between the variability amplitude and rest-frame time lag is observed, in agreement with quasar 
structure functions but extending to AGN several magnitudes fainter than previously studied. The structure function slopes for our fainter AGN sample are slightly shallower 
than those found in quasars studies. An anticorrelation with luminosity is clearly detected, with lower luminosity AGN displaying greater variability amplitudes. We demonstrate 
for the first time that this anticorrelation extends to AGN as faint as $M_{AGN_i}\sim-18.5$, with a slight trend towards shallower slopes at luminosities fainter than 
$M_{AGN_i}\sim-20.2$.

\end{abstract}

\keywords{galaxies: active {\sbond} galaxies: nuclei {\sbond} techniques: photometric}


\section{Introduction} \label{intro}

Luminosity variability is a common feature of quasars (QSOs), and active galactic nuclei (AGN) in general, throughout the electromagnetic spectrum from X-rays to radio 
wavelengths and on timescales from several hours to many years. Various models have been postulated to explain this variability, such as accretion disk instabilities 
\citep[e.g.,][]{rees84,kaw98}, variation of accretion rates \citep[e.g.,][]{li08,zuo12}, supernova explosions or starbursts \citep[e.g.,][]{ter92,kaw98}, and gravitational 
microlensing \citep[e.g.,][]{haw93,alex95}.

Several studies of quasar variability in the optical bands have explored relations between variability amplitude and important parameters such as time lag, luminosity, 
rest-frame wavelength and black hole mass. The amplitude of variability is found to correlate with time lag, increasing until it seems to flatten at longer timescales 
\citep[e.g.,][]{hook94,tre94,cris96,dicle96,vb04,bau09,kel09}.

A number of studies have also found an anticorrelation of the variability with the luminosity of quasars, with more luminous quasars varying less \citep[e.g.,][]
{hook94,tre94,cris96,vb04,wil08,bau09,zuo12}. Furthermore, evidence of an increase of the amplitude of variability with decreasing rest-frame wavelength (bluer) is seen in 
the part of the spectrum ranging from the UV to the near-infrared \citep[e.g.,][]{dicle96,cris97,giv99,hel01,vb04,zuo12}.

In this paper our goal is to extend the study of the variability of QSOs to fainter AGN and investigate whether these trends continue to fainter luminosities. We take an 
approach similar to \citet[hereafter \vbens]{vb04}, who used a sample of $\sim$25,000 quasars drawn from the Sloan Digital Sky Survey \citep[SDSS;][]{sdss} to carry out a 
study of the optical ensemble variability of quasars. 

We compile our sample of morphologically extended, lower luminosity, AGN from the SDSS Seventh Data Release \citep[DR7;][]{dr7}. The broadband imaging photometry can be 
directly compared with spectrophotometry derived from the spectra, taken at a different epoch, for three of the SDSS bands, namely \gba, \rba~and \iba. This provides us 
with data of the magnitude of each object at two epochs and enables the analysis of the ensemble properties of AGN as a function of various parameters on timescales from 
weeks up to 8 years.

The AGN sample is described in \S~\ref{data}. We calculate the contribution of the host galaxy to the AGN component in \S~\ref{decomp}. The definition of the ensemble 
variability is given in \S~\ref{ens}. We show the variability as a function of time lag, luminosity and wavelength in \S~\ref{res} and the results are discussed in \S~\ref{disc}.

Throughout this paper, we assume $\Lambda$CDM cosmology with $\Omega_M=0.27$, $\Omega_\Lambda=0.73$, and $H_0=71~\text{km s$^{-1}$ Mpc$^{-1}$}$.


\section{The AGN Data Set} \label{data}


\subsection{The Sloan Digital Sky Survey} \label{sdss}

The SDSS is a survey that images over 10,000 deg\up{2} in five broad bands and obtains follow-up spectra for roughly a million galaxies and 100,000 quasars. 
The observations are made using a dedicated wide-field 2.5 m telescope \citep{gunn06} located at Apache Point Observatory near Sacramento Peak in Southern New Mexico. 
The images are taken with a 54-CCD camera \citep{gunn98} in drift-scan mode using five filters $u,g,r,i~\text{and}~z$ \citep{fuk96}. The SDSS photometric system is 
calibrated so that the magnitudes are on the AB system \citep{oke83,smi02}. The photometricity and extinction are monitored by a 0.5 m telescope on site \citep{hog01,tuc06}. 
Point source astrometry is accurate to better than 100 mas \citep{pier03}. The magnitudes are corrected for Galactic extinction following \citet{sch98}.

Objects are selected for follow-up spectroscopic observations as candidate galaxies \citep{eis01,stra02}, quasars \citep{rich02} or stars \citep{edr}. The spectra cover from 
3800 {\angs} to 9200 {\angs} at a resolution of $\lambda/\Delta\lambda\simeq2000$. 


\subsection{Sample Selection} \label{sample}

To assemble our sample, we extract the data from the SDSS-DR7 database making use of the SDSS CasJobs site\footnote{\url{http://casjobs.sdss.org/CasJobs}}, 
where an SQL query is submitted specifying the relevant information and constraints. For each object, photometric data from two epochs are obtained, 
one from imaging and one from spectroscopy where the spectroscopic flux is summed over the same band as the imaging photometry (see \S~\ref{var}).

The focus of this work is to study a sample of AGN extending to low luminosities, most likely to be Seyfert galaxies and thus having an extended morphology. We select objects 
contained in the \emph{`View' GALAXY}, defined by having the parameter $obj\_type=3$ \citep[see \S~4.4.6 of][for more details]{edr} and thus implying that they are extended, 
not stars or point-like sources. In addition, we want our galaxies to be spectroscopically classified as having broad emission lines characteristic of type-1 AGN.
We also select a sample of ``normal galaxies'', i.e., resolved objects without broad emission lines, that will act as our control sample used to quantify the photometric noise 
of non-varying galaxies (\S~\ref{var}).

We limit both data sets to have redshifts of $z\lesssim0.84$ in order for the spectra to contain the H$\beta$ and [OIII]$_{\lambda5007}$ emission lines.
These lines may be used in future studies to calculate the mass of the supermassive black hole (SMBH) and $\sigma^*$ of the bulge of the host galaxy, respectively.

Further restrictions are applied regarding the width of the emission lines for the AGN. The SDSS database provides the $\sigma$ measured for a Gaussian fit to the emission lines.
From $\sigma$ we infer the FWHM of the line as follows:
\begin{equation} \label{eq:fwhm}
\text{FWHM(km s$^{-1}$)}=2.35\times\sigma(\angs)\times c(\text{km s$^{-1}$})/\lambda(\angs)~.
\end{equation}

For AGN with $z\lesssim0.40$ where the H$\alpha$ line is also present, the lower limit for the FWHM of both the H$\alpha$ and H$\beta$ lines is set at 
1500 km s$^{-1}$ to ensure robust detection of broad lines. We also impose an upper limit of 10,000 km s$^{-1}$, which corresponds to the Doppler broadening
of gas around a black hole of mass $\sim$$10^9$ M$_\sun$, the expected upper limit for SMBHs. We impose this limit because we want to avoid spurious unphysical 
measurements derived from poor fits to the SDSS spectral features and we do not expect many SMBHs with masses exceeding this limit.

For those objects with $z\gtrsim0.40$ the H$\alpha$ line falls beyond the SDSS spectral coverage range and the FWHM criteria are then only applied to the H$\beta$ line. 
We take this approach assuming that the \emph{lower-z} results may be extrapolated to \emph{higher-z} and because the distribution of FWHM(H$\beta$) for galaxies at 
$z\lesssim0.4$ is very similar to that at $z\gtrsim0.4$. After applying these criteria we have an AGN sample of 5342 sources from the DR7.

The restrictions imposed on the \emph{normal} galaxy sample are extended morphology and spectroscopic classification indicating no broad emission lines. With the same 
redshift cut of $z\lesssim0.84$, the final sample of normal galaxies contains 764,753 sources.

\subsection{Measuring Variability} \label{var}

To measure the variability of our sources, we calculate the difference in magnitude between two epochs. Photometry for one epoch comes from imaging data 
whereas the second epoch is obtained from the follow-up spectroscopic observations, which were taken from several weeks to 8 years later. The \emph{spectro2d} 
pipeline \citep{edr}, used to reduce and calibrate the spectra, also calculates synthetic spectroscopic magnitudes for the \gba, \rba~and \iba~filters 
(the spectra do not cover the entire wavelength ranges of the \uba~and \zba~bands). The magnitude difference is then computed as $\Delta m = m_{ph} - m_{sp}$.

Our photometric data come from the fiber magnitudes, which correspond to the flux contained within an aperture 3{\arcsec} in diameter, the same size as a 
spectroscopic fiber from which the spectroscopic magnitudes are inferred. Unfortunately, the spectrophotometry is calibrated using PSF magnitudes 
\citep[see][]{dr6} and since the PSF includes light that extends beyond the 3{\arcsec} diameter, this produces an offset from the fiber magnitudes 
of roughly 0.35 mag which varies as a function of spectral \sn.

To quantify this offset, the data are divided into \sn~bins for the galaxy and AGN samples such that each bin contains a similar number of objects. For each bin and 
filter, the center and standard deviation of the magnitude difference distributions are determined. 

\begin{figure*} 
\epsscale{1.7}
\plotone{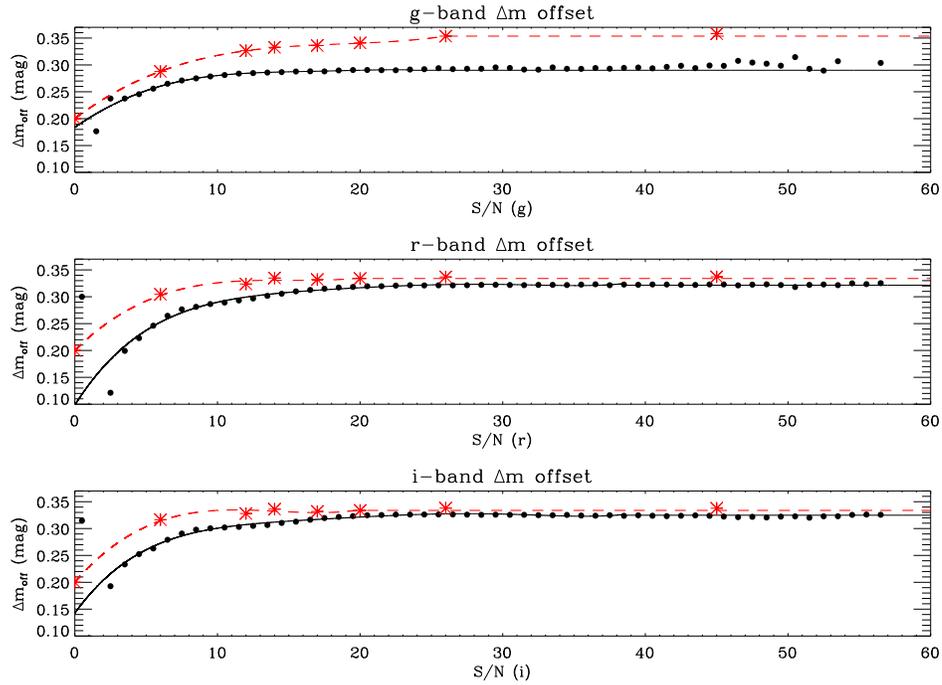}
\caption{Magnitude difference offset vs. spectral \sn~for the galaxy and AGN samples for each of the \gba, \rba~and \iba~filters. The black filled points and 
solid lines correspond to the galaxy sample whereas the red asterisks and dashed lines represent the AGN.\label{fig:offset}}
\end{figure*}

\begin{figure*} 
\epsscale{1.7}
\plotone{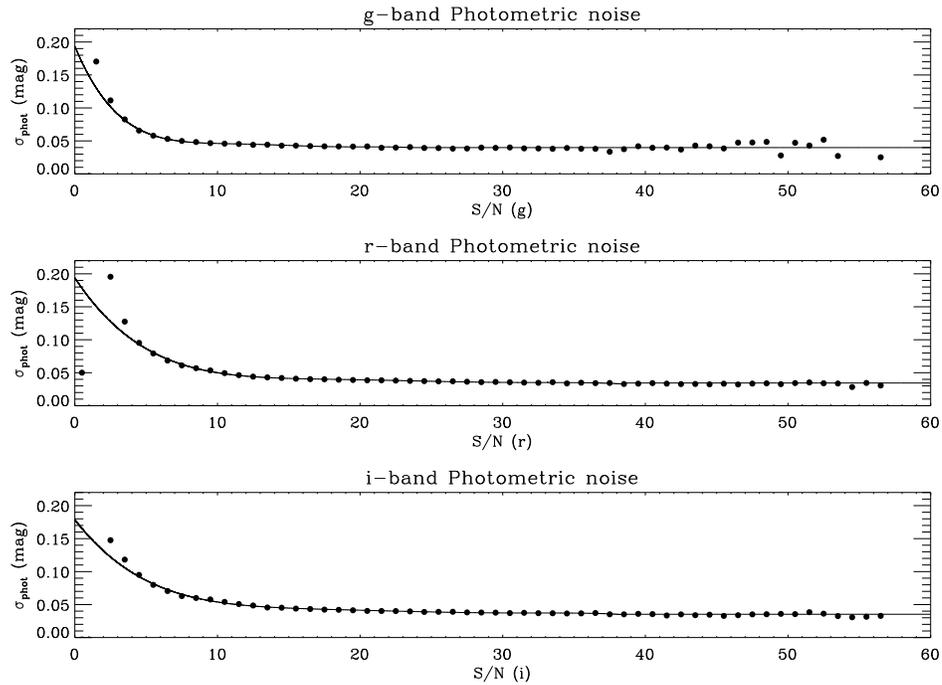}
\caption{Photometric noise vs. spectral \sn~for the galaxy sample for each of the \gba, \rba~and \iba~filters.\label{fig:sigma}}
\end{figure*}

We then analyze the \dm~offset (\dmoff), represented by the values of the centers of the Gaussians, and find that it trends with the spectral signal-to-noise ratio (\sn) 
in such a way that it increases with \sn~and stays constant at high values of \sn. This is shown in Fig.~\ref{fig:offset}, where the black points correspond to the 
\sn~bins of the galaxy sample and the red asterisks to those of the AGN sample. The offset for the AGN sample is larger than the normal galaxies, especially in the 
\gba~band (see Fig.~\ref{fig:offset}). This is likely due to slight differences in morphology between the AGN and galaxy samples (i.e., the AGN are expected to have 
more compact light profiles due to the unresolved and brighter nucleus), which would translate into different aperture corrections. 

\begin{figure*} 
\epsscale{1.6}
\plotone{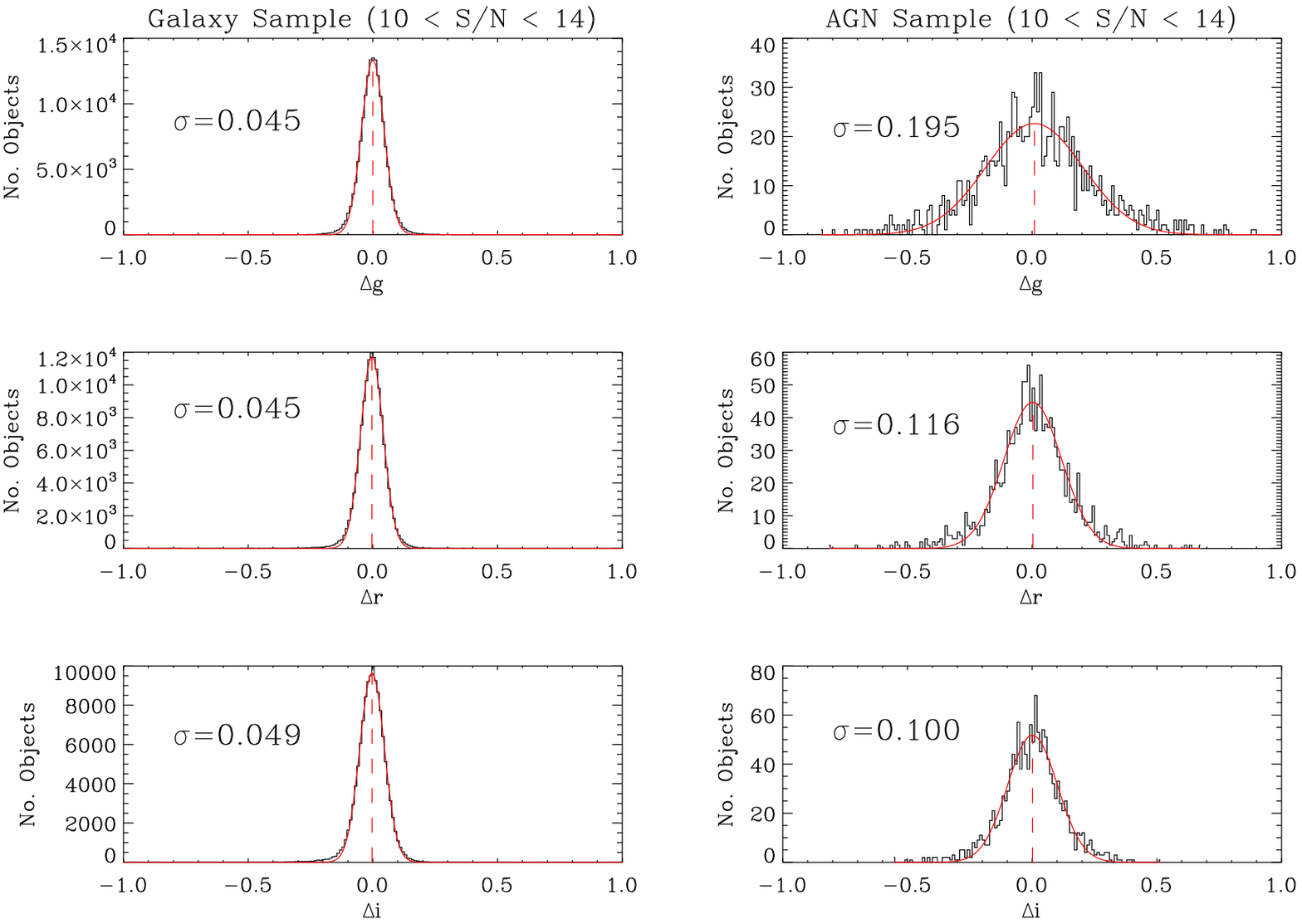}
\caption{Offset corrected magnitude difference histograms in all the three bands for a bin with 10 $<$ \sn~$<$ 14. The red solid lines are the gaussian 
fits to the histograms, whose values of $\sigma$ are given for both the galaxy (\emph{left}) and the AGN (\emph{right}) samples. The vertical red dashed lines 
indicate the center of the distributions.\label{fig:hist}}
\end{figure*}

To model this dependence so that it can be accurately removed, a polynomial function is fitted. For the galaxy sample a 5\up{th} order polynomial is used, whereas for 
the AGN sample a 3\up{rd} order polynomial better describes the data. At high \sn~a constant value is fit. The black solid lines and the dashed red lines represent the 
functions for galaxies and AGN, respectively. We apply this correction to the spectrophotometric magnitudes separately for each sample to remove the systematic offset.

The galaxy sample serves as our control sample of non-varying sources to calculate the photometric noise as a function of \sn. This is determined by dividing the \dm~values of 
the galaxies into \sn~bins and fitting the binned \dm~values with a Gaussian distribution to determine the 1$\sigma$ width (\sigphot; black points in Fig.~\ref{fig:sigma}). 
We model the \sn~dependence with a 5\up{th} order polynomial fit to the points plus a constant at high \sn~(solid line in Fig.~\ref{fig:sigma}).

An example of the \dm~distributions in the \gba, \rba~and \iba~bands for AGN and galaxies with spectral \sn~between 10 and 14 is 
depicted in Fig.~\ref{fig:hist}, where the corresponding $\sigma$ values of the fitted Gaussians (red lines) are given. These values for the galaxy sample are
0.045, 0.045 and 0.049 in the \gba, \rba~and \iba~bands, respectively. For the AGN, the widths are noticeably wider, 0.195, 0.116 and 0.100, for the same bands, 
respectively. The significantly larger magnitude differences found among the AGN sample demonstrates the variable nature of the AGN. 


\section{Spectral Decomposition} \label{decomp}

Before proceeding with the variability analysis of our AGN sample, we need to take into consideration the contamination of the light coming from the host galaxy. 
Since our sources are morphologically extended, the host galaxy's brightness may be comparable or larger than that of the active nucleus. The light from the host galaxy 
is assumed not to vary and thus results in a dampening effect on the observed variability of the AGN, with a larger impact as the host galaxy contribution increases. 

\begin{figure*} 
$\begin{array}{cc}
\includegraphics*[width=0.5 \textwidth]{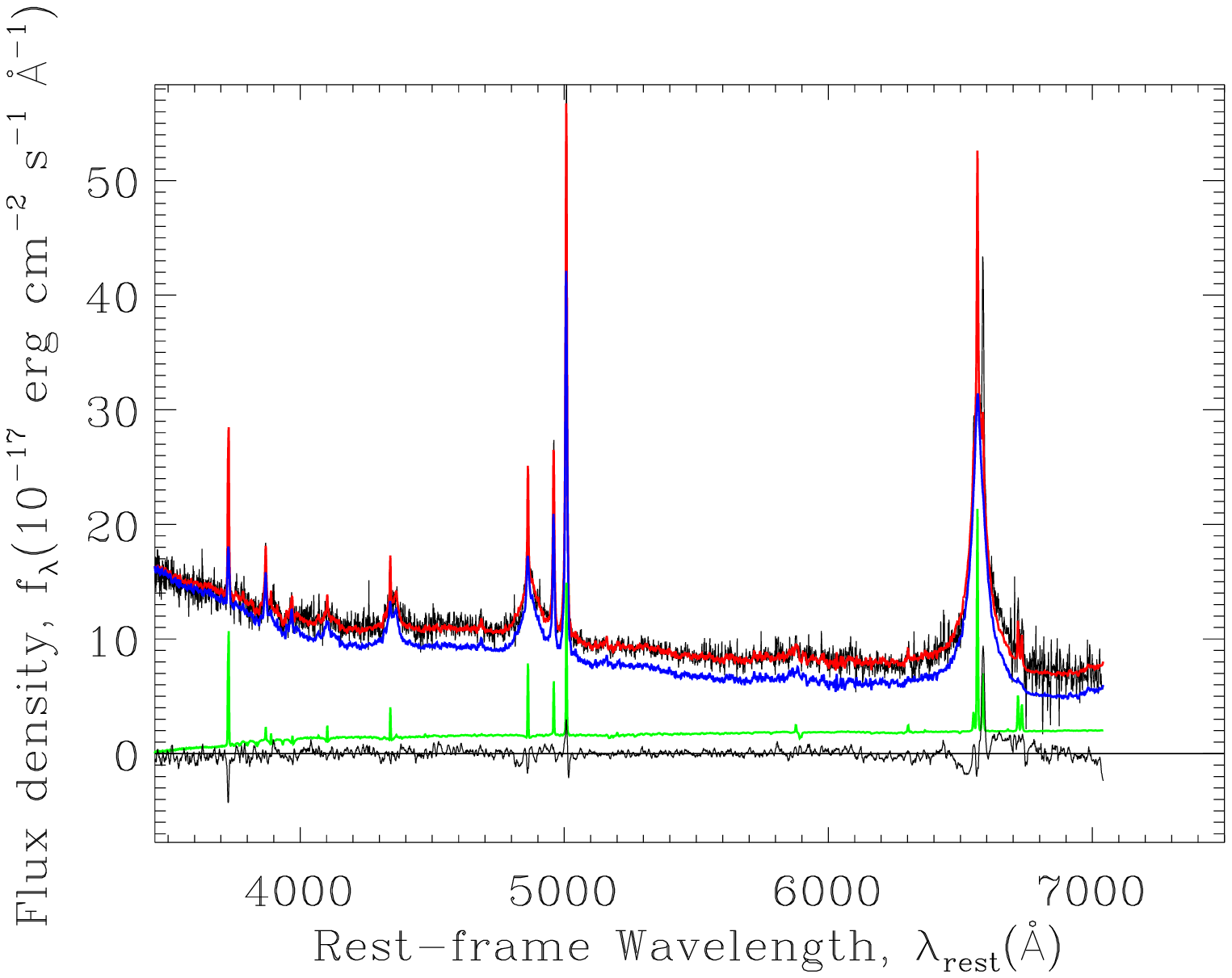}&
\includegraphics*[width=0.5 \textwidth]{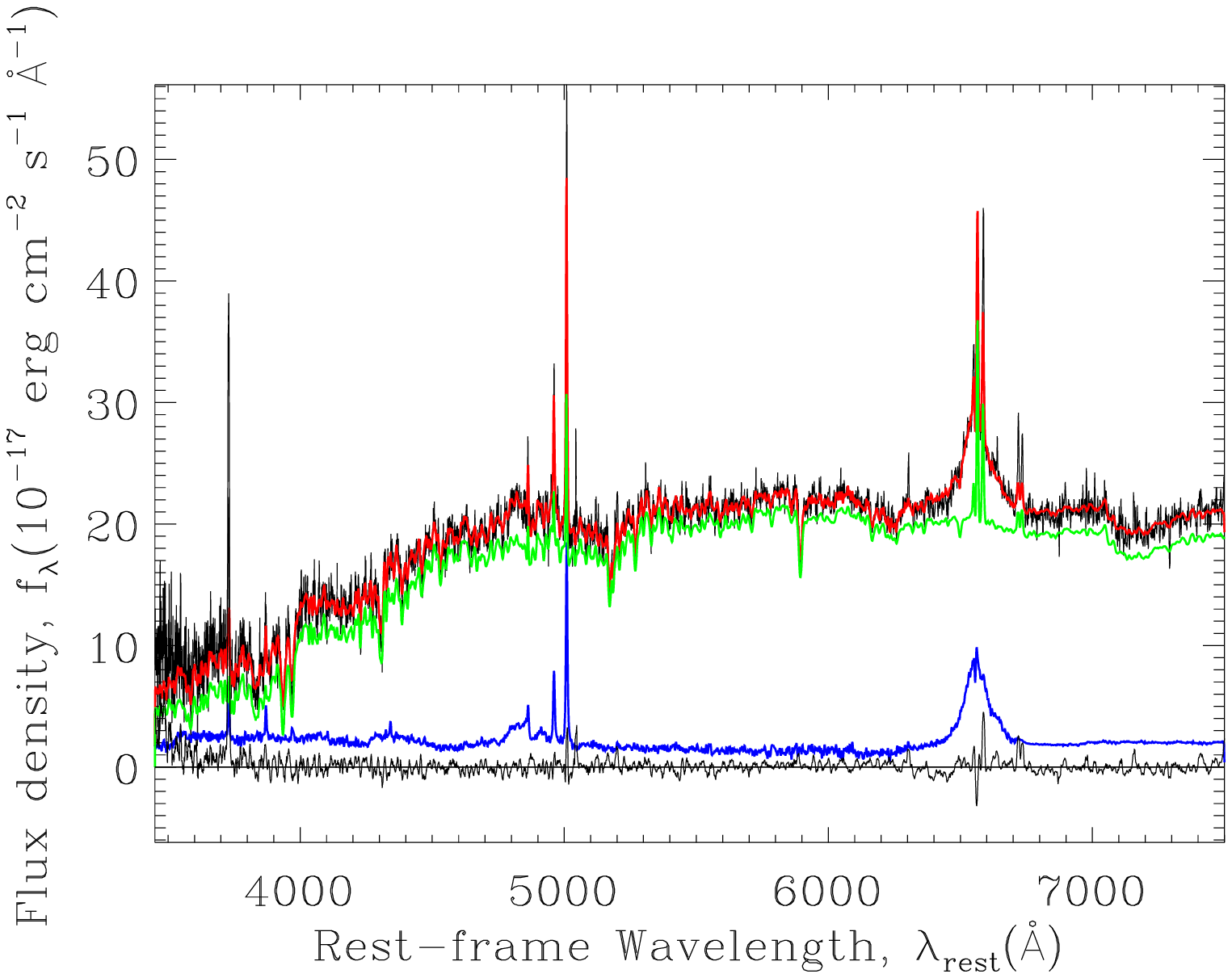}\\
\includegraphics*[width=0.5 \textwidth]{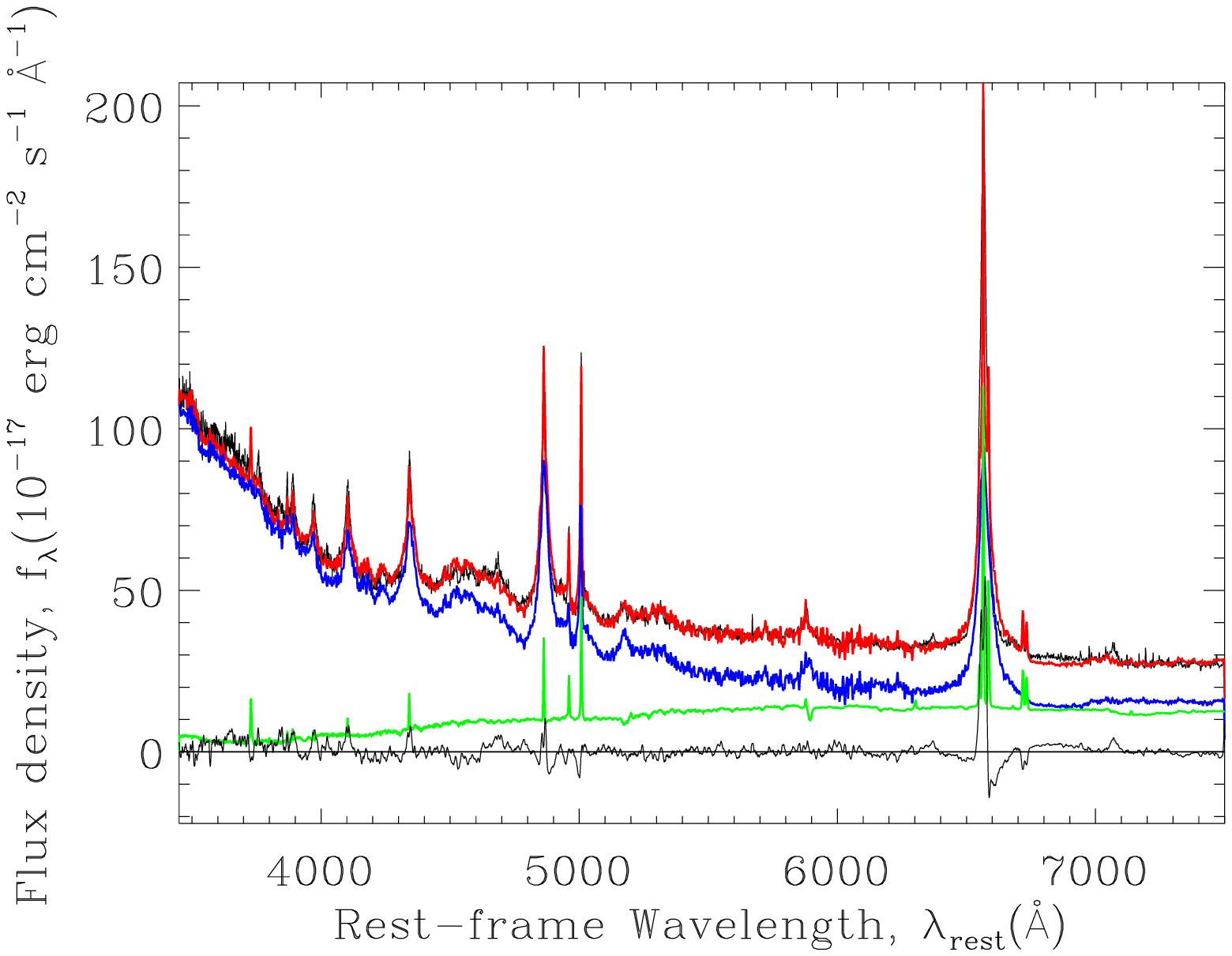}
\end{array}$
\caption{Three examples of AGN/host galaxy spectra reconstructed using qso and galaxy eigenspectra. Each panel shows the original spectrum ($black$), the reconstructed
spectrum ($red$), the AGN component ($blue$), the host galaxy component ($green$), and the residual spectrum smoothed by 7 pixels ($black$; near $f_\lambda=0$).
Top left: the AGN component is dominant. Top right: the host galaxy component is dominant. Bottom: the host galaxy contribution significantly changes along the 
wavelength ($\lambda$) range.\label{fig:recons}}
\end{figure*}

To measure the contribution of the host galaxy component, it is necessary to separate it from the AGN. 
\citet[hereafter \vbdec]{vb06} showed that galaxy or quasar spectra can be reconstructed as linear combinations of eigenspectra described by
\begin{equation} \label{eq:recons}
f_\lambda^R = \sum_{k=1}^m a_k e_k(\lambda)~,
\end{equation}
where $f_\lambda^R$ is the reconstructed flux density as a function of wavelength $\lambda$ and the $a_k$ are the eigencoefficients of the corresponding 
eigenspectra $e_k(\lambda)$. 

We use a modified version of the fitting code of \citet{hao}, with the approach of \vbdec~and employ the sets of eigenspectra described and made available 
by \citet{yipa,yipb} for galaxies and quasars. The quasar eigenspectra are taken from a subset with specific redshift and luminosity ranges. Given that few 
of our AGN sample galaxies are at redshifts above 0.5, we choose the low-redshift bin ``ZBIN 1'', spanning $0.08 < z < 0.53$, and the high-luminosity 
bin for the low-redshift range ``C1'' ($-24 \geq M_i \geq -26$), defined by \citet{yipb}. These templates are luminous enough so that the galaxy contamination in 
the quasar eigenspectra is minimal. Nonetheless, the reconstruction is only reliable up to a redshift limit $z_\text{lim}=0.752$ (see \vbdec). Therefore, we apply 
another cut to our AGN sample, reducing it from 5342 to 5328 objects.

As shown by several authors \citep{con95,con99,yipa}, a relatively small number of eigenspectra can be used to reconstruct a given spectrum because most of the 
information is contained in the first few modes. Following \vbdec, we reconstruct our spectra by means of a combination of five galaxy and ten quasar eigenspectra. 
Three examples are depicted in Fig.~\ref{fig:recons}, where we can see the original spectrum ($black$), the reconstructed spectrum ($red$), the AGN component 
($blue$), the host galaxy component ($green$), and the residual spectrum smoothed by 7 pixels ($black$; near $f_\lambda=0$). The two top panels show galaxies whose 
light is either dominated by the AGN component ($left$) or by the host galaxy component ($right$). The bottom panel illustrates how the contribution of the host 
galaxy to the total galaxy spectrum can vary with wavelength, with the AGN dominating towards the blue end of the spectrum.

To quantify the amount of light that belongs to each of the components, \vbdec~defined the fractional contribution of the host galaxy to the composite spectrum $F_H$, 
hereafter referred to as $\psi$, as the integrated flux densities of the reconstructed quasar and galaxy eigenspectra over the rest-frame wavelength range 
$4160<\lambda<4210\angs$. We take a different approach and calculate $\psi$ as a function of wavelength since the host galaxy contribution varies through the SDSS
bands as shown in Fig.~\ref{fig:recons}. We calculate $\psi$ in wavelength ranges close to the \gba, \rba~and \iba~photometric bands that avoid major galaxy stellar 
absorption lines as well as strong QSO emission lines \citep[see table 30 of][]{edr}. These wavelength regions are used in the following equation to calculate $\psi$,
\begin{equation} \label{eq:psi}
\psi = \frac {\int_{\lambda_1}^{\lambda_2} f_{\lambda,H}^R\,\mathrm{d}\lambda} {\int_{\lambda_1}^{\lambda_2} (f_{\lambda,A}^R + f_{\lambda,H}^R)\,\mathrm{d}\lambda}~,
\end{equation}
where the subscripts $A$ and $H$ denote AGN and host components, respectively. The limits, $\lambda_1$ and $\lambda_2$ (rest-frame), change from filter to filter and with 
redshift to correspond with the \gba, \rba~and \iba~ wavelength regions (observed frame). These regions are typically between $50-150$, $150-250$, and $250-400$ 
{\angs} wide for the \gba, \rba~and \iba~bands, respectively. As an example, at a redshift of $\sim0.25$ the values $[\lambda_1,\lambda_2]$ are $[3800,3900]$, $[4450,4700]$, 
and $[5400,5800]$ {\angs} for the \gba, \rba~and \iba~bands, respectively.

\begin{figure*} 
\epsscale{1.6}
\plotone{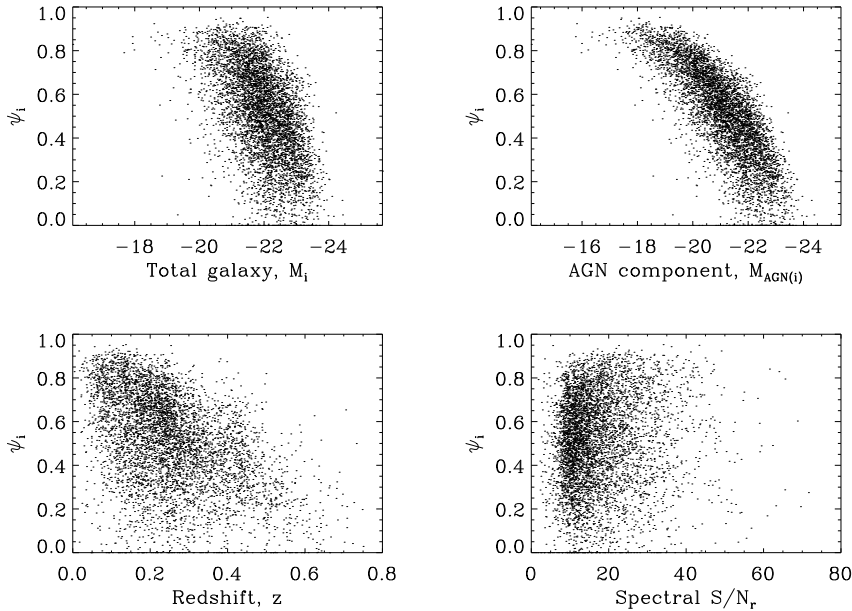}
\caption{Distribution of the AGN sample showing the host-to-galaxy flux fraction in the \iba-band ($\psi_i$) as a function of rest-frame \iba-band absolute magnitude of 
the total galaxy ($M_i$), rest-frame \iba-band absolute magnitude of the AGN component ($M_{AGN_i}$), redshift ($z$) and spectral \sn~in the \rba-band ($\text{\sn}_r$).
\label{fig:param}}
\end{figure*}

Determining $\psi$ is crucial for studying the true variability due to the central engine and correcting for the dampening effect produced by the non-variable host 
galaxy component. $\psi$ yields a representative value for the host galaxy contribution in each filter and this component is assumed to be non-varying. Therefore, 
variability (\dm) that is observed should be corrected by an amount that increases with increasing host galaxy contribution. The true magnitude difference for the 
AGN component alone ($\dmc$) is determined from the observed magnitude change (\dm) and the value of $\psi$ using the following equation:
\begin{eqnarray} \label{eq:dmc}
|\dmc|=2.5\log\left(\frac{f_{2_{AGN}}}{f_{1_{AGN}}}\right)= \nonumber\\
=2.5\left|\log\left(\frac{10^k-\psi}{10^{-k}-\psi}\right)\right|
\end{eqnarray}
where
\begin{displaymath}
k=0.2|\Delta m|~.
\end{displaymath}
Here the quantity in the logarithm represents the fluctuation around an average magnitude value where the contribution from the non-varying host galaxy ($\psi$) has been 
subtracted from the total flux in magnitudes at each of the two epochs ($f_{1_{AGN}}~\text{and}~f_{2_{AGN}}$). The 0.2 multiple in $k$ comes from the usual 0.4 multiple 
used when converting magnitudes into fluxes but divided by 2 since the flux change is taken around the mean value. This value $|\dmc|$ will be used in place of the observed 
$|\Delta m|$ in all subsequent calculations.

We can then determine the absolute magnitude for the AGN component in the rest-frame \iba-band using \emph{k-correct} v4\_2 \citep{kcor} and the calculated $\psi$.
Since the rest-frame absolute magnitude ($M_i$) is derived from the observed photometric magnitude of the entire object, we correct for the contribution of the host 
galaxy component ($\psi$) in order to determine the luminosity of the AGN component. We use $\psi_i$ to correct $M_i$ for our AGN since this is the filter at the longest 
$\lambda$ and best corresponds to the rest-frame \iba-band. Given that $\psi$ is the fraction of the flux of the host galaxy component to the full galaxy, the rest-frame 
\iba-band absolute magnitude of the AGN is determined by
\begin{equation} \label{eq:absmag}
M_{AGN_i} = M_i - 2.5 \log (1-\psi_i)~.
\end{equation}

In Fig.~\ref{fig:param}, we show the distribution of $\psi$ values for our AGN sample as a function of various parameters. We choose the \iba-band to plot $\psi$ as a 
function of rest-frame \iba-band absolute magnitude of the total galaxy, rest-frame \iba-band absolute magnitude of the AGN component alone, redshift and spectral \sn~in 
the \rba-band. The top left panel of the figure shows that galaxies with smaller $\psi$ values are generally brighter. This is likely due to the fact that the AGN dominates 
in these cases and the AGN component adds to the overall brightness, increasing the total luminosity. The top right panel shows that the luminosity of the AGN component, in 
sources where the host galaxy dominates (i.e., larger $\psi$ values), is generally quite faint, reaching absolute magnitudes fainter than $\sim-18$ in some cases. The bottom 
left panel shows that most of the high-$\psi$ sources are at low redshift. This selection effect occurs because the intrinsically fainter sources are not detected at higher 
redshift due to the magnitude limit of the survey. There is no trend with $\psi$ and \sn, indicating that the full range of AGN-to-host galaxy ratios are detected at all 
\sn~values in the survey.


\section{Ensemble Variability} \label{ens}

We compute the ensemble variability, $V$, following that used by \vbens,
\begin{equation} \label{eq:ensvar}
V=\sqrt{\frac{\pi}{2}\langle|\Delta m|\rangle^2-\langle\sigma^2_{phot}\rangle}~,
\end{equation}
where \dm~represents the magnitude difference of each AGN ($\dmc$ from equation (\ref{eq:dmc})), averaged for all the objects within bins of time lag or absolute 
magnitude, \sigphot~is the photometric noise of \dm~as a function of \sn~derived from the control galaxy sample in \S~\ref{var}, whose squared values are averaged 
for all AGN within the mentioned bins, and the factor $\pi$/2 assumes that the distributions of noise and photometric variability are both Gaussian.

The error bars are obtained by applying standard error propagation to equation (\ref{eq:ensvar}),
\begin{equation} \label{eq:errbars}
\sigma(V)={1\over2}V^{-1}\sqrt{\pi^2\langle|\Delta m|\rangle^2~\sigma^2(\langle|\Delta m|\rangle)+\sigma^2(\langle\sigma^2_{phot}\rangle)}~,
\end{equation}
where the $\sigma^2$'s represent the uncertainties in $\langle|\Delta m|\rangle$ and $\langle\sigma^2_{phot}\rangle$. These uncertainties represent the standard error in 
the mean for each bin rather than simply the standard deviation.

When computing the ensemble variability, we have applied further constraints to produce a more robust sample. We include only sources with spectral \sn$_r>4$ so that the 
broad lines are clearly distinguishable, which reduces the sample by $\sim$0.8\%. In order to avoid spurious, non-physical outliers we impose $|\dmc|<2$  ($\sim$0.5\% 
reduction in the sample). Finally we choose $\psi<0.85$ to avoid overcorrection of the AGN \dm~due to the host galaxy contribution ($\sim$4\% reduction). This results in a 
total AGN sample of 5058 galaxies.


\section{Results} \label{res}


\subsection{Variability vs. Time Lag (Structure Function)} \label{tau}

The variability of quasars and AGN can be computed as a function of rest-frame time lag (the so-called ``structure function'') \citep[e.g., \vbens;][]{dicle96}. The time lag 
for each object is determined by subtracting the MJD of the imaging observation date (mjd\sub{\emph{im}}) from that of the spectroscopic data 
(mjd\sub{\emph{sp}}) 
\begin{equation} \label{eq:time}
\text{time-lag}(\Delta t)=|\text{mjd}_{sp}-\text{mjd}_{im}|~.
\end{equation}
The observed time lag depends on the redshift of the galaxy. Therefore, we compute the rest-frame time lag, $\Delta\tau$, as
\begin{equation} \label{eq:tau}
\Delta\tau=\frac{\Delta t}{1+z}~.
\end{equation}

\begin{figure*} 
\epsscale{1.65}
\plotone{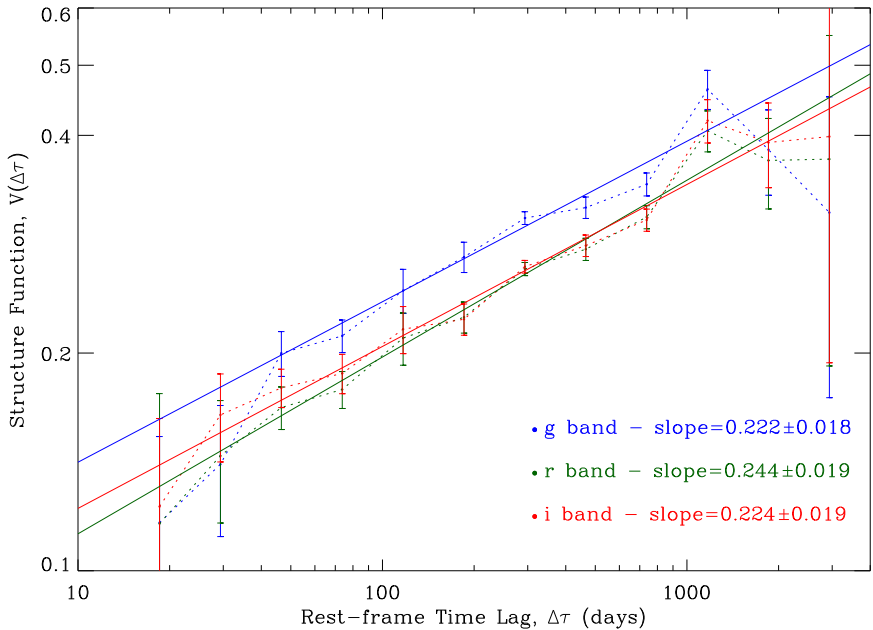} 
\caption{Structure functions of the three bands, color-coded by band. The solid lines are the fits to the single power law. The first two and last two bins have been excluded 
from the fits.\label{fig:lag}}
\end{figure*}

The structure functions for each of the three filters are shown in Fig.~\ref{fig:lag} (blue, green and red represent the \gba, \rba~and \iba~bands, respectively). The 
sample is binned in equal intervals in logarithmic rest-frame time lag, each bin containing tens to hundreds of galaxies, except for the first two and last two bins which 
contain only a few galaxies.

A clear trend is seen with the structure function increasing as a function of rest-frame time lag from 50 to 1000 days, with the amplitude of the change in magnitude becoming 
larger for longer time intervals. This correlation is consistent with what has been found in several studies of QSO variability \citep[e.g.,][]{hook94,tre94,cris96,dicle96,
vb04,bau09,kel09}. We find that the \gba-band (bluest) displays the greatest amplitude of variability compared to the other two bands, in agreement with the anticorrelation with 
wavelength found previously among QSO studies \citep[e.g.,][]{dicle96,cris97,giv99,hel01,vb04,zuo12}.

It is common to fit a power law to the structure function, which appears as a straight line in a log-log scale. The fitted lines are shown in Fig.~\ref{fig:lag} with 
slope values (i.e., power law index values) of $0.222\pm0.018$, $0.244\pm0.019$ and $0.224\pm0.019$ in the \gba, \rba~and \iba~bands, respectively. We do not include the first 
two and last two bins when performing the fit since they contain few objects (10 to 50) compared to the hundreds in the other bins.


\subsection{Variability vs. Absolute Magnitude} \label{absmag}

\begin{figure*} 
\epsscale{1.7}
\plotone{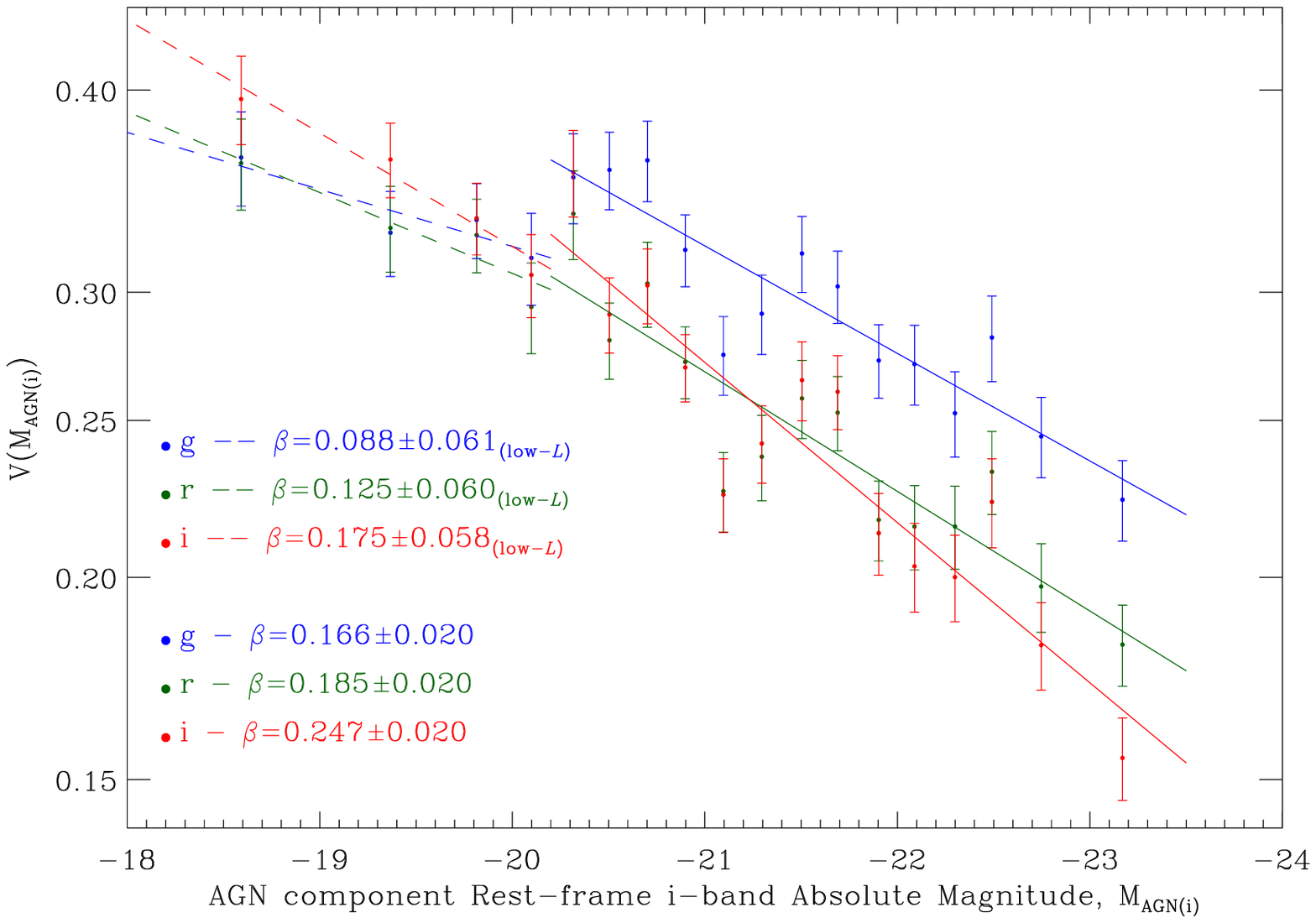} 
\caption{Variability functions vs. rest-frame \iba-band absolute magnitude of the AGN component of the three bands, color-coded by band. The solid and dashed lines are the 
fits to the Poissonian function for the $\log(V)$ vs. $M_{AGN_i}$ representation, for the intermediate-luminosity and low-luminosity bins, respectively. 
\label{fig:miagn}}
\end{figure*}

Many studies have found that fainter AGN and QSOs display greater variability than more luminous sources \citep[e.g.,][]{hook94,tre94,cris96,vb04,wil08,bau09,zuo12}. 
We explore the limits of this trend with our low-luminosity AGN sample. 

Fig.~\ref{fig:miagn} shows the variability functions for each filter (blue, green and red represent the \gba, \rba~and \iba~bands, respectively), where $V(M_{AGN_i})$ 
is plotted in logarithmic scale. The sample is binned such that every bin contains a similar number of galaxies, regardless of the bin size. The center of each bin represents 
the median value of $M_{AGN_i}$. An anticorrelation of the amplitude of variability with AGN luminosity is seen, in agreement with previous studies and extending to the 
faintest absolute magnitudes of our sample, with the least luminous AGN having the greatest variability amplitude.

To compare with other studies, we fit our data to the relation expected from Poissonian models, where the relative variability varies with luminosity as 
$\delta L/L \propto L^{-\beta}$ and $\beta={1\over2}$ in general \citep[e.g.,][]{cid00}. This relationship can be translated into a dependence on absolute luminosity in 
logarithmic form as follows
\begin{equation} \label{eq:poiss}
\log V(M_i) = {\beta\over2.5} M_i + K~,
\end{equation}
where $M_i$ corresponds to $M_{AGN_i}$ and $K$ is a constant.
 
Since a single function does not appear to fit the data well across the entire range of absolute magnitudes for our sample, we have chosen to fit separate functions at 
intermediate luminosities ($-23.5<M_{AGN_i}<-20.2$) and at low luminosities ($-20.2<M_{AGN_i}<-18.5$). We obtain values for the slopes ($\beta$) of $0.166\pm0.020$, 
$0.185\pm0.020$ and $0.247\pm0.020$ for the \gba, \rba~and \iba~bands, respectively, at intermediate luminosities and shallower slopes of $0.088\pm0.061$, $0.125\pm0.060$ 
and $0.175\pm0.058$ for the low-luminosity end of the distribution. The amplitude of variability is higher for the \gba-band (bluest) data at intermediate luminosities, 
consistent with previous studies where sources are found to be more variable at bluer wavelengths and as also seen in the structure function. This difference among bands 
disappears at low luminosities, where the three bands overlap primarily due to a decrease in variability for the g-band at the low-luminosity end. This drop does 
not appear to be due to any significant change in the distribution of redshifts, inclusion of spectral emission lines in the band, or variability correction due to host 
galaxy contamination. It is therefore unclear what is the physical or selection effect that may be causing this decrease. We also find that the g-band slope is 
shallower compared to the redder bands at both intermediate and low luminosities and discuss the possible reasons for this in the following section.


\section{Discussion and Conclusions} \label{disc}

\begin{figure*} 
\epsscale{1.7}
\plotone{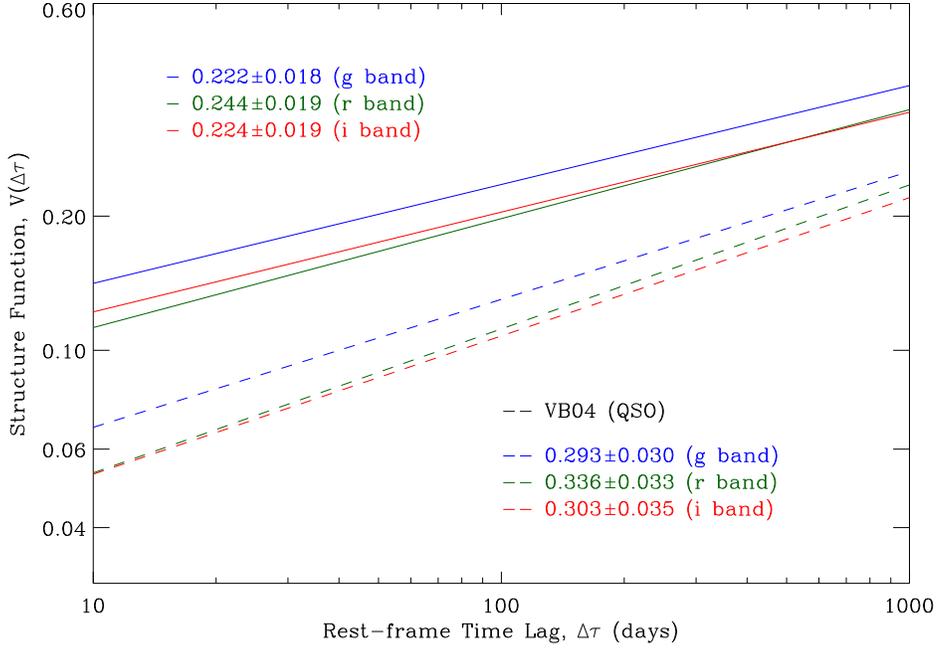} 
\caption{Comparison of our fits of the structure functions to single power laws ($solid~lines$) to those from \vbens~($dashed~lines$), color-coded by band. The values of 
the corresponding slopes are shown.\label{fig:lag_comp}}
\end{figure*}

We have analyzed the dependence of AGN variability amplitude on various parameters with the goal of probing the limits of previously identified trends at lower luminosities. 
On timescales from 50 to 1000 days, we find a positive correlation between rest-frame time lag and variability amplitude (the structure function). We compare the structure 
functions for our AGN sample, which extends to $M_{AGN_i}\sim-18.5$, to those for QSOs presented in \vbens~(Fig.~\ref{fig:lag_comp}). The amplitude of variability 
is significantly greater in all bands and at all timescales for our sample compared to the QSO sample and we discuss this dependency on luminosity later in this section. 
Although different at only the $\sim1.5\sigma$ significance, the slopes of our structure functions are all shallower than the corresponding QSO structure functions in the 
same bands. 

\begin{deluxetable}{lcccr}
\tabletypesize{\footnotesize}
\tablecaption{Comparison of Structure Function Slopes for QSOs and AGN\label{tbl:tau}}
\tablewidth{0pt}
\tablehead{
\colhead{Sample} & \colhead{Band} & \colhead{SF Slope} & \colhead{Normalization} & \colhead{Authors}
}
\startdata
SDSS AGN & \gba & $0.222\pm0.018$ & No & This work \\
SDSS AGN & \rba & $0.244\pm0.019$ & No & This work \\
SDSS AGN & \iba & $0.224\pm0.019$ & No & This work \\
SDSS QSO & \gba & $0.293\pm0.030$ & No & \vbens \\
SDSS QSO & \rba & $0.336\pm0.033$ & No & \vbens \\
SDSS QSO & \iba & $0.303\pm0.035$ & No & \vbens \\
SDSS QSO & \gba\rba\iba & $0.246\pm0.008$ & Yes\tablenotemark{a} & \vbens \\
SDSS-S82 QSO & \gba & $0.479$ & No & \citet{wil08} \\
SDSS-S82 QSO & \rba & $0.486$ & No & \citet{wil08} \\
SDSS-S82 QSO & \iba & $0.436$ & No & \citet{wil08} \\
QUEST2 QSO & $R$ & $0.357\pm0.014$ & No & \citet{bau09} \\
QUEST2 QSO & $R$ & $0.392\pm0.022$ & Yes\tablenotemark{b} & \citet{bau09} \\
QUEST2 QSO & $R$ & $0.432\pm0.024$\tablenotemark{c} & Yes\tablenotemark{b} & \citet{bau09} \\
\enddata
\tablenotetext{a}{Isolate dependence on luminosity by binning the other parameters to calculate $V(M_i)$ for each bin and then normalize the data points from the $gri$ filters 
altogether}
\tablenotetext{b}{Isolate dependence on luminosity by binning the other parameters to calculate $V(M_i)$ for each bin and then normalize to the set with the best statistics}
\tablenotetext{c}{High-luminosity bins}
\end{deluxetable}

\begin{figure*} 
\epsscale{1.8}
\plotone{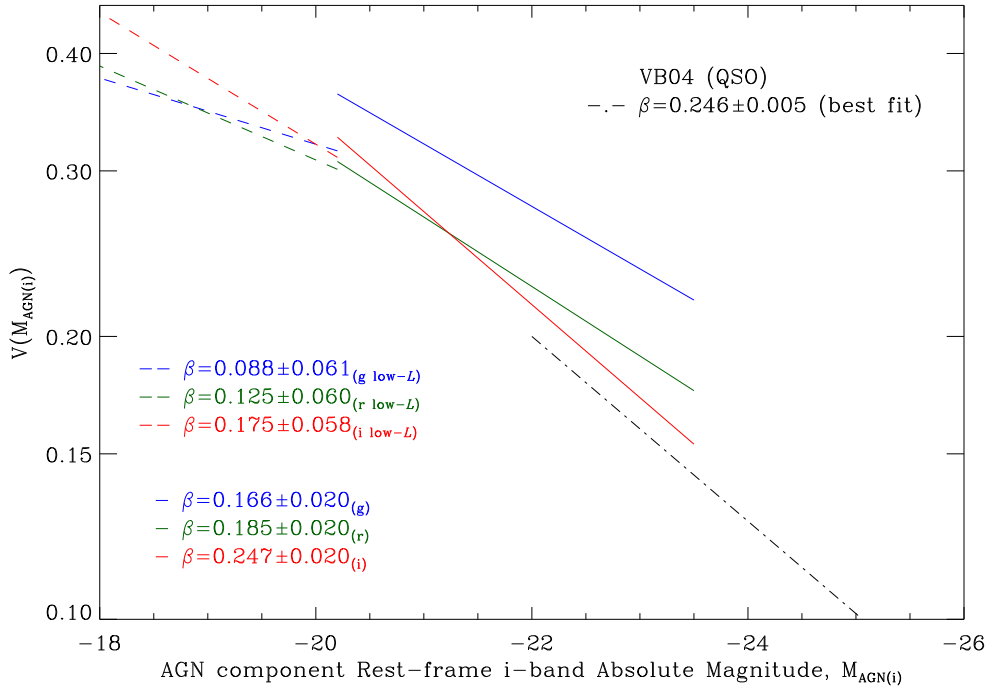} 
\caption{Comparison of our fits (color-coded by band) of the variability functions vs. rest-frame \iba-band absolute magnitude ($solid~lines$ for \emph{intermediate-L} 
and $dashed~lines$ for \emph{low-L}) to the scaled data points from \vbens~for the best-fit (\emph{dash-dotted line}). The values of the corresponding slopes are shown.
\label{fig:miagn_comp}}
\end{figure*}

\begin{deluxetable}{lcccr}
\tabletypesize{\footnotesize}
\tablecaption{Comparison of Variability vs. Luminosity Slopes for QSOs and AGN\label{tbl:miagn}}
\tablewidth{0pt}
\tablehead{
\colhead{Sample} & \colhead{Band} & \colhead{SF Slope} & \colhead{Comments} & \colhead{Authors}
}
\startdata
SDSS AGN & \gba & $0.166\pm0.020$ & Intermediate-\emph{L} & This work \\
SDSS AGN & \rba & $0.185\pm0.020$ & Intermediate-\emph{L} & This work \\
SDSS AGN & \iba & $0.247\pm0.020$ & Intermediate-\emph{L} & This work \\
SDSS QSO & \gba\rba\iba & $0.246\pm0.005$ & Normalized & \vbens \\
QUEST2 QSO & $R$ & $0.205\pm0.002$ & High-\emph{L} & \citet{bau09} \\
SDSS-S82 QSO & \gba\rba\iba & $0.223\pm0.075$ & Median\tablenotemark{a} & \citet{zuo12} \\
SGP QSO & $B$ & $0.5$ & Poissonian model & \citet{cid00} \\
\enddata
\tablenotetext{a}{Median value of the slopes for the three filters and for the different multiparameter binned subsamples}
\end{deluxetable}

Other QSO variability studies such as \citet{wil08} and \citet{bau09} also measure steeper structure function slopes for QSOs (see Table~\ref{tbl:tau}). 
\citet{bau09} found that the normalized structure function for their higher luminosity QSOs was slightly steeper than that found for their entire QSO sample, which is 
consistent with our finding of shallower slopes for lower luminosity sources. However, they attribute this to incompleteness among their lower-luminosity sources which they 
argue do not sample the full range of $\Delta m$ values.

We test for incompleteness in our sample by computing the range of $\Delta m$ values as a function of absolute magnitude. We divide the AGN into several absolute magnitude 
bins and determine the standard deviation ($\sigma$) of the $\Delta m$ in each bin. The standard deviation of the $\Delta m$ distribution is $\sigma=0.227$ in the \gba-band 
for sources with $-22<M_{AGN_i}<-21$. At fainter absolute magnitudes ($-20<M_{AGN_i}<-19$) $\sigma=0.254$ (\gba-band). For sources fainter than $-19$, $\sigma=0.260$ 
(\gba-band). The $\sigma$ values remain roughly the same at all magnitudes and do not appear to decrease at faint absolute magnitudes, as might be expected if the 
full range of possible values is not being sampled. Therefore, we find that incompleteness in the $\Delta m$ range for our AGN does not appear significant to 
$M_{AGN_i}\sim-18.5$ and thus would not impact the slope of the measured structure function.

\citet{kaw98} and \citet{haw02} calculate theoretical structure function slopes using models of disk instabilities, starburst and microlensing with values ranging $0.41-0.49$, 
$0.74-0.90$, and $0.23-0.31$, respectively. Although our slopes are more consistent with the microlensing model, this is an unlikely source of variability for these relatively 
low-redshift AGN. Disk instabilities and changes in the accretion rate are more likely to be the source of variability for the majority of AGN as discussed below and in \vbens.

We observe a clear trend between variability amplitude and absolute magnitude as shown in Fig.~\ref{fig:miagn}. Simple Poissonian models predict a slope with $\beta=0.5$ 
\citep{cid00}, which is much larger and inconsistent with our slopes and those of previous QSO studies. Fig.~\ref{fig:miagn_comp} shows our slopes compared to that of 
QSOs presented in \vbens. In the luminosity range where our sample overlaps their sample ($M_{AGN_i}\sim-22~\text{to} -23.5$), our slopes are in close agreement, particularly 
among the \iba-band data, and continue to magnitudes of $\sim-20.2$. Our intermediate-luminosity slopes are also in good agreement with several other QSO studies 
(Table~\ref{tbl:miagn}).

Based on the standard accretion disk model proposed by \citet{sha73} and following \citet{li08}, \citet{zuo12} calculate model predictions of the variability amplitude 
caused by changes in accretion rate. They find that accretion rate changes of 20\% produce variability amplitude changes that reproduce the tendencies of correlations between 
variability amplitude and luminosity. These models appear the most promising at explaining the qualitative dependency of AGN/QSO variability on absolute magnitude, although 
the models produce much flatter slopes than the observations reveal.

We find that the slope of the variability vs. absolute magnitude is an increasing function with wavelength. This may be due to several combined effects. As previously 
noted, the variability amplitude of AGN has been found to be greatest at bluer rest-$\lambda$ \citep[e.g., \vbens;][]{zuo12}. In addition, spectroscopic studies have revealed 
that AGN are bluer in their brightest phase \citep[e.g., \vbens;][]{dicle96,tre01} and that brighter AGN are generally bluer in color \citep[e.g., \vbens;][]{dicle96}. 
These characteristics combined could result in sources in the brighter, bluer bins of our sample having higher variability amplitudes compared to the sources in the fainter 
and/or redder bins, producing a shallower anticorrelation for the bluer data. The slopes in all three bands become slightly shallower at lower luminosities ($M_{AGN_i}>-20.2$) 
but not significantly above the errors of the fits. Nonetheless, we generally find that the anticorrelation between variability amplitude and absolute magnitude continues to 
the faintest absolute magnitudes of our sample.

In conclusion, the relationships previously observed for QSOs between variability amplitude, rest-frame time lag and absolute magnitude have been found to continue to lower 
luminosity AGN as faint as $M_{AGN_i}\sim-18.5$. We find a strong correlation of variability with rest-frame time lag and an anticorrelation with wavelength since the bluer 
\gba-band exhibits a larger variability amplitude than the redder bands in our sample. We find evidence for shallower structure function slopes for our fainter AGN when 
compared to brighter QSO samples which does not appear to be due to incompleteness. The slope of the anticorrelation between variability and absolute magnitude found for 
QSOs continues through our intermediate-luminosity sources to $M_{AGN_i}\sim-20.2$. At the faintest end of our distribution, the anticorrelation continues but with a slight 
trend towards shallower slopes.


\acknowledgments

We gratefully acknowledge contributions by Tyler Desjardins and Daniel Vanden Berk on an earlier version of this work and who provided us with very helpful comments. 
We thank Lei Hao for providing us with her code to fit the quasar and galaxy eigenspectra.

Funding for the SDSS and SDSS-II has been provided by the Alfred P. Sloan Foundation, the Participating Institutions, the National Science Foundation, the U.S. Department 
of Energy, the National Aeronautics and Space Administration, the Japanese Monbukagakusho, and the Max Planck Society, and the Higher Education Funding Council for England. 
The SDSS Web site is http://www.sdss.org/.

The SDSS is managed by the Astrophysical Research Consortium (ARC) for the Participating Institutions. The Participating Institutions are the American Museum of Natural 
History, Astrophysical Institute Potsdam, University of Basel, University of Cambridge, Case Western Reserve University, The University of Chicago, Drexel University, 
Fermilab, the Institute for Advanced Study, the Japan Participation Group, The Johns Hopkins University, the Joint Institute for Nuclear Astrophysics, the Kavli Institute 
for Particle Astrophysics and Cosmology, the Korean Scientist Group, the Chinese Academy of Sciences (LAMOST), Los Alamos National Laboratory, the Max-Planck-Institute for 
Astronomy (MPIA), the Max-Planck-Institute for Astrophysics (MPA), New Mexico State University, Ohio State University, University of Pittsburgh, University of Portsmouth, 
Princeton University, the United States Naval Observatory, and the University of Washington.


\clearpage



\end{document}